\documentclass[lettersize,journal]{IEEEtran}

\usepackage{amsmath,amsfonts,amssymb}
\usepackage{array}
\usepackage[caption=false,font=normalsize,labelfont=sf,textfont=sf]{subfig}
\usepackage{textcomp}
\usepackage{stfloats}
\usepackage{url}
\usepackage{verbatim}
\usepackage{graphicx}
\usepackage{cite}

\usepackage{algorithm}
\usepackage{algpseudocode}

\usepackage{tabularray}
\UseTblrLibrary{booktabs}
\usepackage{bm}

\usepackage{siunitx}

\usepackage{hyperref}
\usepackage{orcidlink}

\newcommand*{\dif}{\mathop{}\!\mathrm{d}}

\begin{document}

\title{Weak-Fluctuation-Induced Clutter Covariance and Subspace Structure in Single-Snapshot FDA-MIMO GPR}

\author{Yisu Yan$^{\orcidlink{0009-0008-4022-3619}}$,~\IEEEmembership{Graduate Student Member,~IEEE,} Jifeng Guo$^{\orcidlink{0000-0002-9710-0045}}$ 

\thanks{Yisu Yan is with School of Astronautics, Harbin Institute of Technology, Heilongjiang, China
(e-mail: \href{mailto:23B918085@stu.hit.edu.cn}{23B918085@stu.hit.edu.cn}).}

\thanks{Jifeng Guo is with School of Astronautics, Harbin Institute of Technology, Heilongjiang, China (e-mail: \href{mailto:guojifeng@hit.edu.cn}{guojifeng@hit.edu.cn}).}

\thanks{This work has been submitted to the IEEE for possible publication. Copyright may be transferred without notice, after which this version may no longer be accessible.}

}

\maketitle

\begin{abstract}
	Weak constitutive fluctuations in dispersive subsurface media can induce distributed clutter that reshapes the observation structure of ground-penetrating radar (GPR). This paper analyzes this effect for single-snapshot frequency-diverse array multiple-input multiple-output GPR. Focusing on medium-induced clutter, rather than on general target--clutter joint modeling, it establishes a statistical propagation chain from Cole--Cole parameter perturbations to electromagnetic contrast, first-order Born channel snapshots, clutter covariance, and subspace descriptors. A medium-aware snapshot model and a covariance propagation framework are then derived to characterize how constitutive uncertainty alters observation-domain spectral structure under a local weak-fluctuation regime. Numerical experiments verify the consistency of the proposed propagation relation under the adopted first-order Born and constitutive-linearization approximations. Within the tested setting, medium-induced clutter reshapes the eigenspectrum and changes target--clutter overlap metrics. Spatial correlation length and background-scene variation act as consistently strong structural drivers, while the FDA frequency increment also produces measurable changes in the normalized covariance geometry.
\end{abstract}


\section{Introduction}\label{sec:intro}

\IEEEPARstart{I}{n} many GPR applications, small-scale non-target fluctuations in soil, weathered layers, or lunar regolith generate distributed backscattering. This background component can alter the apparent channel-domain structure and complicate subsequent interpretation. Similar effects have been reported in landmine-oriented GPR \cite{takahashi2010InfluenceSoilInhomogeneity,takahashi2012ModelingGPRClutter}, planetary shallow-subsurface sensing \cite{feng2022ShallowRegolithStructure,zhang2025SubsoilStructureChangE6}, and shallow geological surveying \cite{salinasnaval2018GPRClutterAmplitude}. These observations indicate that weak distributed medium fluctuations are a real and relevant source of clutter.

This issue is especially important in single-snapshot FDA-MIMO-GPR, where the observation is formed by one space--frequency measurement under near-field subsurface propagation. The resulting channel-domain structure depends on the array, the frequency-indexed transmit configuration, and the background propagation kernel, which may include interface-transmission effects when they are represented in the prescribed Green response.

Relevant studies can be reviewed from three complementary perspectives. The first concerns clutter statistics, covariance processing, and adaptive detection in FDA/FDA-MIMO radar. Existing studies have analyzed clutter rank and degrees of freedom for frequency-diverse waveforms and FDA-MIMO configurations \cite{liu2016ClutterRanksFrequency,wang2022ClutterRankAnalysis}. They have also developed clutter-suppression and waveform-design methods based on multi-waveform processing, tensor-based STAP, compressed sensing, and parameter optimization \cite{WEN2019280,wang2022RangeAmbiguousClutterSuppression,sun2024SpaceTimeRange,jia2026FDAMIMORadarParameter}. More recent work continues to formulate FDA-MIMO target detection mainly through detector design and covariance manipulation under assumed statistical backgrounds \cite{li2026AdaptiveTargetDetection}. This body of work provides useful statistical descriptors, including clutter rank, clutter subspace, and target--clutter separability. However, the clutter field is usually prescribed at the signal-processing level, and its formation from subsurface constitutive fluctuations is not explicitly modeled.

A second relevant perspective is provided by GPR and MIMO-GPR forward modeling, where the subsurface propagation mechanism is treated more explicitly. Early FDA-MIMO-GPR formulations considered near-subsurface target detection, but generally relied on a homogeneous background with constant propagation parameters and a scalar reflection coefficient \cite{liu2018DetectionSubsurfaceTarget}. More recent multichannel GPR studies employ multiview, multistatic, and multifrequency forward models that can account for air--soil transmission, refracted propagation paths, and medium-dependent constitutive parameters \cite{masoodi2024MultiviewMultistaticVs,gennarelli2026EffectEquivalentPermittivity}. Related planetary radar observations also show that shallow-subsurface responses may be governed by irregular regolith structures rather than by idealized layered interfaces alone \cite{feng2022ShallowRegolithStructure,zhang2025SubsoilStructureChangE6}. In parallel, much of the GPR clutter literature has focused on post-formation mitigation, including signal-domain suppression, spectral filtering, and clutter-distribution normalization \cite{kumar2025EnhancingSubsurfaceExploration,oliveira2021GPRClutterReflection,worthmann2021ClutterDistributionsTomographic}. These studies provide physically grounded forward models and effective mitigation tools, but they do not directly characterize how medium-induced clutter forms a covariance structure in a single-snapshot FDA-MIMO observation.

A further body of literature concerns constitutive dispersion and parameter uncertainty in subsurface media. Classical dielectric-dispersion models, including Debye- and Cole--Cole-type descriptions, show that complex permittivity is intrinsically frequency dependent \cite{cole1941DispersionAbsorptionDielectrics,auty1952DielectricPropertiesIce,holm2020TimeDomainCharacterization}. This frequency dependence affects radar interpretation and inversion. For example, Qin \emph{et al.} \cite{qin2023FullwaveformInversionGroundpenetrating} showed that neglecting attenuation and velocity dispersion can significantly degrade the recovery of conductive and scattering-dominated anomalies in GPR full-waveform inversion. Uncertainty-aware studies have likewise emphasized the need to quantify parameter variability in subsurface interpretation \cite{xie2021GPRUncertaintyModelling}. Material and planetary studies further indicate that dielectric behavior varies with composition, frequency, temperature, and porosity \cite{strangway1972ElectricalPropertiesLunar,olhoeft1973ElectricalPropertiesLunar,olhoeft1974ElectricalPropertiesLunar,brouet2016CharacterizationPermittivityControlled,elshafie2013DielectricHardnessMeasurements}. Numerical modeling has also moved toward representing arbitrary complex dielectric properties in full-wave simulations \cite{majchrowska2021ModellingArbitraryComplex}. These results support the need to retain an explicit constitutive dependence when modeling clutter induced by subsurface media.

Overall, the existing literature provides statistical descriptors for FDA/FDA-MIMO clutter, physically grounded GPR forward models, and constitutive descriptions of dispersive media. What remains less developed is a statistical propagation framework that connects weak Cole--Cole parameter perturbations to electromagnetic contrast, single-snapshot FDA-MIMO channel responses, clutter covariance, and the resulting spectral and subspace structure.

This paper addresses that gap in a restricted setting. The focus is on clutter induced by weak, distributed, non-target constitutive fluctuations in complex dispersive media under a single-snapshot FDA-MIMO-GPR observation. The paper does \emph{not} attempt to model all clutter sources, strong interface echoes, or general target--clutter interactions. Instead, it isolates one medium-induced clutter component that is physically meaningful and analytically tractable. Target steering vectors are introduced only as geometric probes for overlap with the dominant clutter subspace.

The remainder of this paper is organized as follows. Section~\ref{sec:snapshot_model} introduces the weak distributed clutter snapshot model in a complex dispersive medium and derives the statistical propagation from constitutive perturbations to electromagnetic contrast and channel snapshots. Section~\ref{sec:clutter_analysis} constructs the medium-induced and total clutter covariances and analyzes their structural implications. Section~\ref{sec:experiments} presents numerical validation and structural diagnostics. Section~\ref{sec:conclusion} concludes the paper.

\section{Medium-Aware Snapshot Model for FDA-MIMO-GPR}\label{sec:snapshot_model}

This section establishes a medium-aware single-snapshot model for FDA-MIMO-GPR in a dispersive subsurface environment. The model is intended for weak distributed non-target medium fluctuations around a prescribed background state. It is derived under a background-perturbation formulation, a first-order Born approximation, a first-order constitutive linearization, and a scalarized channel-domain representation. These assumptions define the scope of the subsequent covariance and subspace analysis.

\subsection{Cole--Cole Representation of the Background Medium}\label{subsec:background_medium}

The subsurface medium is described by the Cole--Cole constitutive model. Let
\begin{equation}
	\bm{\mu}(\bm{x})
	=
	\big[
		\varepsilon_{\infty}(\bm{x}),
		\Delta\varepsilon(\bm{x}),
		\tau(\bm{x}),
		\alpha(\bm{x}),
		\sigma(\bm{x})
		\big]^{\top}
	\label{eq:mu_vector}
\end{equation}
denote the local parameter vector, where $\varepsilon_{\infty}$ is the high-frequency permittivity, $\Delta\varepsilon$ is the relaxation strength, $\tau$ is the characteristic relaxation time, $\alpha$ is the broadening parameter, and $\sigma$ is the conductivity. The corresponding complex permittivity is
\begin{equation}
	\varepsilon_c(\omega,\bm{x})
	=
	\varepsilon_0
	\left[
		\varepsilon_{\infty}(\bm{x})
		+
		\frac{\Delta\varepsilon(\bm{x})}{1+\big(j\omega\tau(\bm{x})\big)^{1-\alpha(\bm{x})}}
		-
		\frac{j\sigma(\bm{x})}{\omega\varepsilon_0}
		\right]
	\label{eq:cole_cole_local}
\end{equation}

To separate large-scale propagation from local constitutive fluctuation, the medium is decomposed as
\begin{equation}
	\varepsilon_c(\omega,\bm{x})
	=
	\varepsilon_b(\omega,\bm{x})
	+
	\delta\varepsilon(\omega,\bm{x})
	\label{eq:eps_decomposition}
\end{equation}
where $\varepsilon_b$ denotes the background medium and $\delta\varepsilon$ denotes the local deviation. This is a \emph{modeling decomposition} that specifies a reference background about which the propagation and scattering operators are constructed. The resulting model therefore characterizes clutter induced by constitutive perturbations around a prescribed dispersive background $\varepsilon_b(\omega,\bm{x})$, in the form of Eq.~\eqref{eq:cole_cole_local}.

\subsection{Background Green Response and Kernel Generation}\label{subsec:kernel_generation}

Under the $e^{j\omega t}$ convention and the nonmagnetic assumption $\mu(\bm{x})\approx\mu_0$, the background electric field satisfies
\begin{equation}
	\nabla\times\nabla\times\bm{E}(\bm{x},\omega)
	-
	\omega^2\mu_0\varepsilon_b(\omega,\bm{x})\bm{E}(\bm{x},\omega)
	=
	-j\omega\mu_0\bm{J}(\bm{x},\omega)
	\label{eq:background_wave_equation}
\end{equation}
Define the background Maxwell operator by
\begin{equation}
	\mathcal{L}_b(\omega)\bm{E}
	:=
	\nabla\times\nabla\times\bm{E}
	-
	\omega^2\mu_0\varepsilon_b(\omega,\bm{x})\bm{E}
	\label{eq:background_operator}
\end{equation}
Let $\bm{G}_b(\bm{x},\bm{x}';\omega)$ denote the associated Green tensor, satisfying
\begin{equation}
	\mathcal{L}_b(\omega)\bm{G}_b(\bm{x},\bm{x}';\omega)
	=
	\bm{I}\,\delta(\bm{x}-\bm{x}')
	\label{eq:green_tensor_background}
\end{equation}

For the $n$th transmit element located at $\bm{s}_n$, let $\bm{J}_{t,n}(\bm{x};\omega_n)$ denote the impressed source current. The corresponding background incident field is
\begin{equation}
	\bm{E}^{\mathrm{inc}}_n(\bm{x},\omega_n)
	=
	-j\omega_n\mu_0
	\int
	\bm{G}_b(\bm{x},\bm{x}';\omega_n)\,
	\bm{J}_{t,n}(\bm{x}';\omega_n)\,
	\dif\bm{x}'.
	\label{eq:bg_incident_field_general}
\end{equation}
For a localized transmit element, this field is written in the compact channel form
\begin{equation}
	\bm{E}^{\mathrm{inc}}_n(\bm{x},\omega_n)
	=
	\bm{p}_{t,n}(\omega_n)\,
	G_t(\bm{x},\bm{s}_n;\omega_n)
	\label{eq:Gt_definition}
\end{equation}
where $\bm{p}_{t,n}(\omega_n)$ collects source amplitude and polarization factors, and $G_t(\bm{x},\bm{s}_n;\omega_n)$ is the effective transmit kernel generated from the background Green response.

For the $m$th receive element located at $\bm{r}_m$, let $\mathcal{M}_m[\cdot]$ denote the linear receive functional. Applied to a field $\bm{u}(\bm{r},\omega_n)$, it gives the scalar channel output
\begin{equation}
	\mathcal{M}_m[\bm{u}]
	=
	\int
	\bm{w}_m^{\mathrm H}(\bm{r};\omega_n)\,
	\bm{u}(\bm{r},\omega_n)\,
	\dif\bm{r}
	\label{eq:measurement_functional}
\end{equation}
where $\bm{w}_m$ is the receive test function. The corresponding receive kernel from a subsurface point $\bm{x}$ to the $m$th channel is
\begin{equation}
	G_r(\bm{r}_m,\bm{x};\omega_n)
	:=
	\mathcal{M}_m\!\left[
		\bm{G}_b(\cdot,\bm{x};\omega_n)
		\right]
	\label{eq:Gr_definition}
\end{equation}

To obtain a tractable channel-domain model, a scalarized single-polarization representation is adopted. Specifically, the transmit excitation, the receive measurement, and the background response are projected onto the effective polarization states of the array channels. Under this reduction, the vector Green response is represented by the scalar propagation kernels $G_t$ and $G_r$ in \eqref{eq:Gt_definition} and \eqref{eq:Gr_definition}. The kernels are therefore defined with respect to the prescribed background medium. They retain the dispersive propagation effects, and possible interface effects, that are included in $\bm{G}_b$. They do \emph{not} include higher-order feedback caused by repeated interaction between the local perturbation and the scattered field.

\subsection{Single-Snapshot Forward Model}\label{subsec:forward_model}

Consider an FDA-MIMO array with transmit frequencies
\begin{equation}
	\omega_n=\omega_0+(n-1)\Delta\omega,
	\quad n=1,\ldots,N
	\label{eq:fda_frequency}
\end{equation}
where $\omega_0$ is the first transmit angular frequency and $\Delta\omega$ is the frequency increment. Under the \emph{first-order Born approximation}, the total field inside the scattering integral is replaced by the background incident field. The echo observed at the $(m,n)$ channel pair is then modeled as
\begin{equation}
	\begin{aligned}
		y_{mn}(\omega_n)
		= &
		\int_D
		G_r(\bm{r}_m,\bm{x};\omega_n)\,
		\xi(\bm{x},\omega_n)\,
		G_t(\bm{x},\bm{s}_n;\omega_n)\,
		\dif\bm{x} \\
		  & +
		n_{mn}(\omega_n)
	\end{aligned}
	\label{eq:refined_snapshot_model}
\end{equation}
where $\xi(\bm{x},\omega)$ is the electromagnetic contrast relative to the background medium, defined as
\begin{equation}
	\xi(\bm{x},\omega)
	:=
	\frac{\delta\varepsilon(\omega,\bm{x})}{\varepsilon_b(\omega,\bm{x})}
	\label{eq:contrast_definition}
\end{equation}
and $n_{mn}(\omega_n)$ denotes additive noise and residual modeling error.

The first-order Born approximation is adopted to describe weak distributed non-target fluctuations embedded in a dispersive background medium. Under this approximation, the total field inside the scattering integral is replaced by the background incident field, and the observation retains only the leading-order contribution of the constitutive perturbation. The resulting model is intended for \emph{weak distributed} clutter formation.

\subsection{Cole--Cole Parameter Perturbation Mapping}\label{subsec:parameter_mapping}

Let $\bm{\mu}_b(\bm{x})$ denote the background Cole--Cole parameter vector and define the local parameter perturbation by
\begin{equation}
	\delta\bm{\mu}(\bm{x})
	=
	\bm{\mu}(\bm{x})-\bm{\mu}_b(\bm{x})
	\label{eq:delta_mu_definition}
\end{equation}

The perturbation vector $\delta\bm{\mu}$ is introduced as a local linearization variable around the background state $\bm{\mu}_b(\bm{x})$. The analysis is therefore restricted to sufficiently small perturbations for which the first-order expansion remains accurate and the perturbed constitutive response stays within a physically meaningful neighborhood of the background parameter set.

Since the complex permittivity is a deterministic function of the Cole--Cole parameters, one may write
\begin{equation}
	\varepsilon_c(\omega,\bm{x})
	=
	F_{\mathrm{CC}}\big(\omega;\bm{\mu}(\bm{x})\big)
	\label{eq:Fcc_definition}
\end{equation}
For sufficiently small parameter perturbations, \emph{first-order} expansion at $\bm{\mu}_b(\bm{x})$ yields
\begin{equation}
	\delta\varepsilon(\omega,\bm{x})
	\approx
	\sum_{q=1}^{5}
	\frac{\partial F_{\mathrm{CC}}\big(\omega;\bm{\mu}_b(\bm{x})\big)}{\partial \mu_q}
	\,\delta\mu_q(\bm{x})
	\label{eq:delta_epsilon_first_order}
\end{equation}
Substituting \eqref{eq:delta_epsilon_first_order} into \eqref{eq:contrast_definition} gives
\begin{equation}
	\xi(\bm{x},\omega)
	\approx
	\sum_{q=1}^{5}\psi_q(\omega,\bm{x})\,\delta\mu_q(\bm{x})
	\label{eq:contrast_psi_form}
\end{equation}
where
\begin{equation}
	\psi_q(\omega,\bm{x})
	:=
	\frac{1}{\varepsilon_b(\omega,\bm{x})}
	\frac{\partial F_{\mathrm{CC}}\big(\omega;\bm{\mu}_b(\bm{x})\big)}{\partial \mu_q}
	\label{eq:psi_definition}
\end{equation}

The first-order constitutive linearization introduced in Eq.~\eqref{eq:delta_epsilon_first_order} and \eqref{eq:contrast_definition} leads to a tractable observation operator. The resulting error is governed by neglected second-order and higher-order terms in $\delta\bm{\mu}$. Its effect becomes visible when the perturbation amplitude is large or when the constitutive mapping is strongly nonlinear over the perturbation range. This limit will be discussed through numerical experiments in Sec.~\ref{subsec:exp_linearization_boundary}.

Substituting \eqref{eq:contrast_psi_form} into \eqref{eq:refined_snapshot_model} yields
\begin{equation}
	\begin{gathered}
		y_{mn}(\omega_n)
		\approx               \\
		\sum_{q=1}^{5}
		\int_D
		G_r(\bm{r}_m,\bm{x};\omega_n)\,
		\psi_q(\omega_n,\bm{x})\,
		\delta\mu_q(\bm{x})\,
		G_t(\bm{x},\bm{s}_n;\omega_n)\,
		\dif\bm{x}               \\
		+
		n_{mn}(\omega_n)
	\end{gathered}
	\label{eq:y_mn_mu_continuous}
\end{equation}

Partition the investigation domain $D$ into $P$ cells $\{D_p\}_{p=1}^{P}$, and approximate the field quantities as locally constant within each cell. Let $\bm{x}_p$ denote the representative point and $\Delta V_p$ the cell volume. Then \eqref{eq:y_mn_mu_continuous} becomes
\begin{equation}
	y_{mn}(\omega_n)
	\approx
	\sum_{p=1}^{P}\sum_{q=1}^{5}
	K_{mn}^{(q)}(\bm{x}_p;\omega_n)\,
	\delta\mu_q(\bm{x}_p)\,
	\Delta V_p
	+
	n_{mn}(\omega_n)
	\label{eq:y_mn_discrete}
\end{equation}
with
\begin{equation}
	K_{mn}^{(q)}(\bm{x};\omega_n)
	:=
	G_r(\bm{r}_m,\bm{x};\omega_n)\,
	\psi_q(\omega_n,\bm{x})\,
	G_t(\bm{x},\bm{s}_n;\omega_n)
	\label{eq:kernel_definition}
\end{equation}

The accuracy of this discretization depends on whether the cell size is fine enough to resolve the spatial variation of the propagation kernels, the sensitivity functions, and the perturbation field.

Stack all channel observations into
$
	\bm{y}
	\in
	\mathbb{C}^{MN\times 1}
$
and collect the discretized parameter perturbations into
$
	\delta\bm{\mu}
	\in
	\mathbb{R}^{5P\times 1}
$.
Then the snapshot model is written in the compact linear form
\begin{equation}
	\bm{y}
	\approx
	\bm{A}\,\delta\bm{\mu}
	+
	\bm{n},
	\label{eq:final_snapshot_model}
\end{equation}
where $\bm{A}\in\mathbb{C}^{MN\times 5P}$ is the forward matrix induced by the background propagation kernels and the Cole--Cole sensitivity functions.

Equations \eqref{eq:refined_snapshot_model}--\eqref{eq:final_snapshot_model} are therefore accurate up to four modeling layers: background-medium decomposition, first-order Born scattering, first-order constitutive linearization, and spatial discretization. These approximations define the applicability range of the proposed snapshot model and also explain the error terms examined later in Sec.~\ref{sec:experiments}.

\section{Medium-Induced Clutter Analysis}\label{sec:clutter_analysis}

Based on the snapshot model in \eqref{eq:final_snapshot_model}, this section analyzes clutter induced by weak distributed non-target constitutive fluctuations in a complex dispersive background medium. The focus is on how such fluctuations generate observation-domain covariance, reshape the clutter spectrum, and alter dominant-subspace geometry in single-snapshot FDA-MIMO-GPR.

The present analysis considers weak distributed medium fluctuations around a prescribed background state. The perturbation field is modeled as a zero-mean random field,

\begin{equation}
	\mathbb{E}\!\left[\delta\bm{\mu}\right]=\bm{0}
	\label{eq:zero_mean_delta_mu}
\end{equation}
so that the induced clutter component also satisfies
\begin{equation}
	\mathbb{E}\!\left[\bm{c}\right]=\bm{0}
	\label{eq:zero_mean_c}
\end{equation}
The additive noise term is assumed to be zero-mean and uncorrelated with the medium perturbation,
\begin{equation}
	\mathbb{E}\!\left[\bm{n}\right]=\bm{0},
	\quad
	\mathbb{E}\!\left[\delta\bm{\mu}\,\bm{n}^{\mathrm H}\right]=\bm{0}
	\label{eq:noise_uncorrelated_assumption}
\end{equation}
Under these assumptions, the second-order moments introduced below coincide with covariance matrices in the strict sense.

\subsection{Medium-Induced Clutter Covariance}\label{subsec:clutter_covariance}

The observation vector is written as Eq.~\eqref{eq:final_snapshot_model}. The random observation component driven by the medium perturbation is defined as
\begin{equation}
	\bm c
	:=
	\bm A\,\delta\bm\mu
	\label{eq:clutter_vector}
\end{equation}
Accordingly,
\begin{equation}
	\bm y
	=
	\bm c+\bm n
	\label{eq:y_clutter_noise}
\end{equation}

Let
\begin{equation}
	\bm R_\mu
	:=
	\mathbb E\!\left[\delta\bm\mu\,\delta\bm\mu^{\top}\right]
	\in\mathbb{R}^{5P\times 5P}
	\label{eq:Rmu_definition}
\end{equation}
denote the covariance of the parameter perturbation field. The corresponding medium-induced clutter covariance is
\begin{equation}
	\bm R_c
	:=
	\mathbb E\!\left[\bm c\bm c^{\mathrm H}\right]
	=
	\bm A\,\bm R_\mu\,\bm A^{\mathrm H}
	\label{eq:Rc_basic}
\end{equation}
Therefore, the observation covariance satisfies
\begin{equation}
	\bm R_y
	\approx
	\bm R_c+\bm R_n
	\label{eq:Ry_Rc_Rn}
\end{equation}
where $\bm R_n:=\mathbb E[\bm n\bm n^{\mathrm H}]$.

Equation \eqref{eq:Rc_basic} is the basic covariance-propagation relation, which shows that medium-induced clutter is the image of the constitutive-parameter covariance under the observation operator $\bm A$.

\subsection{Parameter-Channel Decomposition}\label{subsec:parameter_channel_decomposition}

The stacked perturbation vector is partitioned as
\begin{equation}
	\delta\bm\mu
	=
	\begin{bmatrix}
		\delta\bm\mu_1^{\top} &
		\delta\bm\mu_2^{\top} &
		\delta\bm\mu_3^{\top} &
		\delta\bm\mu_4^{\top} &
		\delta\bm\mu_5^{\top}
	\end{bmatrix}^{\top}
	\label{eq:delta_mu_stack}
\end{equation}
where $\delta\bm\mu_q\in\mathbb{R}^{P\times 1}$ is the discretized perturbation field of the $q$th Cole--Cole parameter. Consistently, the forward matrix is written as
\begin{equation}
	\bm A
	=
	\begin{bmatrix}
		\bm A_1 & \bm A_2 & \bm A_3 & \bm A_4 & \bm A_5
	\end{bmatrix}
	\label{eq:A_block}
\end{equation}
with $\bm A_q\in\mathbb{C}^{MN\times P}$ denoting the channel matrix associated with the $q$th parameter component.

The covariance matrix $\bm R_\mu$ admits the block form
\begin{equation}
	\bm R_\mu
	=
	\begin{bmatrix}
		\bm R_{\mu,11} & \bm R_{\mu,12} & \cdots & \bm R_{\mu,15} \\
		\bm R_{\mu,21} & \bm R_{\mu,22} & \cdots & \bm R_{\mu,25} \\
		\vdots         & \vdots         & \ddots & \vdots         \\
		\bm R_{\mu,51} & \bm R_{\mu,52} & \cdots & \bm R_{\mu,55}
	\end{bmatrix}
	\label{eq:Rmu_block}
\end{equation}
where
\begin{equation}
	\bm R_{\mu,qq'}
	:=
	\mathbb E\!\left[\delta\bm\mu_q\,\delta\bm\mu_{q'}^{\top}\right]
	\in\mathbb{R}^{P\times P}
	\label{eq:Rmu_qq_definition}
\end{equation}
The diagonal blocks describe spatial covariance within each parameter channel, while the off-diagonal blocks describe cross-correlation between different Cole--Cole parameters.

Substituting \eqref{eq:A_block} and \eqref{eq:Rmu_block} into \eqref{eq:Rc_basic} gives
\begin{equation}
	\bm R_c
	=
	\sum_{q=1}^{5}\sum_{q'=1}^{5}
	\bm A_q\,\bm R_{\mu,qq'}\,\bm A_{q'}^{\mathrm H}
	\label{eq:Rc_sum_qq}
\end{equation}
This decomposition shows that the clutter covariance is formed jointly by all parameter channels. When $\bm R_{\mu,qq'}\neq\bm 0$ for $q\neq q'$, cross-channel coupling generates additional observation-domain structure through the terms
\begin{equation}
	\bm A_q\,\bm R_{\mu,qq'}\,\bm A_{q'}^{\mathrm H}
	\label{eq:Rc_cross_term}
\end{equation}

\subsection{Basic Structural Properties}\label{subsec:basic_structural_properties}

Since $\bm R_c$ is a covariance matrix, it is Hermitian and positive semidefinite:
\begin{equation}
	\bm R_c=\bm R_c^{\mathrm H},
	\qquad
	\bm R_c\succeq 0
	\label{eq:Rc_psd}
\end{equation}
Hence, all eigenvalues of $\bm R_c$ are nonnegative real numbers, and its spectral structure is well defined.

Moreover, by rank inequalities,
\begin{equation}
	\operatorname{rank}(\bm R_c)
	=
	\operatorname{rank}(\bm A\bm R_\mu\bm A^{\mathrm H})
	\le
	\min\!\left\{
	\operatorname{rank}(\bm A),
	\operatorname{rank}(\bm R_\mu)
	\right\}
	\label{eq:rank_bound_basic}
\end{equation}
Since $\bm R_\mu\in\mathbb{R}^{5P\times 5P}$,
\begin{equation}
	\operatorname{rank}(\bm R_c)\le 5P
	\label{eq:rank_bound_5P}
\end{equation}
Therefore, the observation-domain clutter degrees of freedom are jointly limited by the statistical degrees of freedom of the medium perturbation field and by the distinguishability imposed by the forward mapping $\bm A$.

\subsection{Modal Interpretation and Spectral Spreading}\label{subsec:modal_interpretation}

Let the eigendecomposition of $\bm R_\mu$ be
\begin{equation}
	\bm R_\mu
	=
	\bm U_\mu\bm\Lambda_\mu\bm U_\mu^{\top}
	\label{eq:Rmu_eig}
\end{equation}
where
\begin{equation}
	\bm\Lambda_\mu
	=
	\operatorname{diag}(\lambda_1,\lambda_2,\ldots,\lambda_{5P}),
	\quad
	\lambda_\ell\ge 0
	\label{eq:Lambda_mu}
\end{equation}
Let $\bm u_\ell$ denote the $\ell$th eigenvector of $\bm R_\mu$, and define its observation-domain image by
\begin{equation}
	\bm b_\ell
	:=
	\bm A\,\bm u_\ell
	\label{eq:b_l_definition}
\end{equation}
Then \eqref{eq:Rc_basic} becomes
\begin{equation}
	\bm R_c
	=
	\sum_{\ell=1}^{5P}
	\lambda_\ell\,\bm b_\ell\bm b_\ell^{\mathrm H}
	\label{eq:Rc_modal_sum}
\end{equation}

Each statistical mode of the constitutive perturbation field is mapped by $\bm A$ into an observation-domain clutter mode. The final covariance is the weighted superposition of all such modal outer products.

Equation \eqref{eq:Rc_modal_sum} also provides a useful interpretation of spectral spreading. A broader clutter spectrum is more likely to arise when the parameter-field covariance is not concentrated on only a few dominant modes and when the mapped observation-domain vectors $\{\bm b_\ell\}$ occupy sufficiently diverse directions. This interpretation is qualitative. It is used to explain the subsequent numerical observations rather than as a formal sufficient-condition theorem.

Let the eigendecomposition of $\bm R_c$ be
\begin{equation}
	\bm R_c
	=
	\bm U_c\bm\Lambda_c\bm U_c^{\mathrm H}
	\label{eq:Rc_eig}
\end{equation}
with
\begin{equation}
	\bm\Lambda_c
	=
	\operatorname{diag}(\nu_1,\nu_2,\ldots,\nu_{MN}),
	\quad
	\nu_1\ge \nu_2\ge\cdots\ge \nu_{MN}\ge 0
	\label{eq:Lambda_c}
\end{equation}
Define the normalized eigenvalues by
\begin{equation}
	\bar{\nu}_i
	:=
	\frac{\nu_i}{\sum_{k=1}^{MN}\nu_k}
	\label{eq:normalized_eigs}
\end{equation}
The effective rank of $\bm R_c$ is then defined as
\begin{equation}
	r_{\mathrm{eff}}(\bm R_c)
	:=
	\exp\!\left(
	-\sum_{i=1}^{MN}\bar{\nu}_i\ln\bar{\nu}_i
	\right)
	\label{eq:effective_rank}
\end{equation}
A larger value of $r_{\mathrm{eff}}(\bm R_c)$ indicates a more distributed spectrum and a less concentrated clutter energy profile.

\subsection{Effective Clutter Subspace Dimension}\label{subsec:effective_dimension}

For a prescribed energy level $\rho\in(0,1)$, define the effective clutter-subspace dimension as
\begin{equation}
	p_\rho
	:=
	\min\left\{
	p:
	\frac{\sum_{i=1}^{p}\nu_i}{\sum_{i=1}^{MN}\nu_i}
	\ge
	\rho
	\right\}
	\label{eq:p_rho_definition}
\end{equation}
This quantity measures how many dominant eigen-directions are required to capture a prescribed proportion of clutter energy. It complements the effective rank: the former is threshold-based, while the latter summarizes global spectral dispersion.

\subsection{Target--Clutter Separability}\label{subsec:target_clutter_separability}

Let $\bm a_t\in\mathbb{C}^{MN\times 1}$ denote the steering vector of a target of interest, and let
\begin{equation}
	\bm U_p
	=
	\begin{bmatrix}
		\bm u_{c,1} & \bm u_{c,2} & \cdots & \bm u_{c,p}
	\end{bmatrix}
	\label{eq:Up_definition}
\end{equation}
contain the first $p$ eigenvectors of $\bm R_c$. The target energy captured by the dominant clutter subspace is defined as
\begin{equation}
	\eta(p)
	:=
	\frac{\|\bm U_p^{\mathrm H}\bm a_t\|_2^2}{\|\bm a_t\|_2^2}
	\label{eq:eta_definition}
\end{equation}
A larger $\eta(p)$ indicates stronger overlap between the target steering direction and the dominant clutter subspace. A complementary separability measure is
\begin{equation}
	\gamma(p)
	:=
	1-\eta(p),
	\label{eq:gamma_definition}
\end{equation}
which represents the fraction of target energy lying outside the dominant clutter subspace. Larger values of $\gamma(p)$ correspond to better target--clutter separability.

Together, $\bm R_c$, $r_{\mathrm{eff}}(\bm R_c)$, $p_\rho$, $\eta(p)$, and $\gamma(p)$ provide a compact description of medium-induced clutter formation under the present weak-fluctuation regime. These quantities are used in the next section as structural diagnostics rather than as universal descriptors for arbitrary strong-scattering clutter.

\section{Experimental Results and Analysis}\label{sec:experiments}

This section validates the local weak-fluctuation propagation chain
\begin{equation*}
	\delta\bm{\mu}
	\rightarrow
	\delta\varepsilon
	\rightarrow
	\xi
	\rightarrow
	\bm{c}
	\rightarrow
	\bm{R}_c
\end{equation*}
for single-snapshot FDA-MIMO-GPR, and then examines how background media, FDA frequency offsets, spatial correlation, inter-parameter coupling, target geometry, and interpretation boundaries reshape the clutter covariance
\begin{equation*}
	\bm{R}_c=\bm{A}\bm{R}_{\mu}\bm{A}^{\mathrm H}.
\end{equation*}

\subsection{Experimental Setup}\label{subsec:exp_setup}

Six background scenarios are used. The physically interpretable cases $S_1$--$S_4$ correspond to lunar regolith, dry basalt/lava, pure ice, and moist sandy-loam soil, respectively, whereas $S_{\mathrm{syn}}$ and $S_{\mathrm{balance}}$ serve as diagnostic synthetic references. Scenario $S_4$ is included as a physically referenced lossy-soil stress case for the proposed weak-fluctuation chain.

All experiments use the same $8\times 8$ FDA-MIMO geometry, the same $x$--$z$ discretization with $P=525$ cells, and the same scalarized background propagation kernel. The standardized perturbation model
\begin{equation*}
	\begin{gathered}
		\delta\bm{\mu}
		=
		(\bm{D}_{\mu}\otimes\bm{I}_P)
		\delta\widetilde{\bm{\mu}}
		\\
		\bm D_\mu
		=
		\operatorname{diag}
		\left(
		d_{\varepsilon_\infty},
		d_{\Delta\varepsilon},
		d_{\tau},
		d_{\alpha},
		d_{\sigma}
		\right)
	\end{gathered}
\end{equation*}
is used throughout. The diagonal entries of $\bm D_\mu$ define the physical scale of one unit standardized perturbation in each Cole--Cole parameter channel.

Unless otherwise stated, the standardized random field adopts $\ell_x=\qty{0.15}{m}$, $\rho_c=0.3$, and $w=[1,1,1,1,1]$, the Monte Carlo sample size is 2000, and the random seed is 20260405. In the downstream structural experiments, the analysis amplitude is clipped to $s_{\mu}=1$ whenever the validity scan does not identify a smaller admissible scale.

\begin{table*}[!t]
	\centering
	\caption{Background scenarios and Cole--Cole parameters used in the experiments.}
	\label{tab:exp_scene_setup}
	\begin{talltblr}[
		label = none,
		entry = none,
		remark{Note} = {$S_1$--$S_3$ are physically interpretable baselines used to cover weakly dispersive lunar regolith, weakly dispersive dry rock, and Debye-limit ice backgrounds. $S_4$ is a moist sandy-loam diagnostic case derived from the reduced soil model in \cite{Loewer2017UltraBroadbandSoils,Miller2004RadarDetectionLandmines}, used as a lossy shallow-soil stress case. $S_{\mathrm{syn}}$ is used for chain diagnostics, whereas $S_{\mathrm{balance}}$ is used for the parameter-coupling scans.}
		]{
		width=\textwidth,
		colspec={X[0.8]X[2.0]X[1]X[1]X[1]X[1]X[1]},
		row{1}={font=\bfseries},
		cells={font=\footnotesize,valign=m}
		}
		\toprule
		Scenario               & Name                                                                                                    & $\varepsilon_{\infty}$ & $\Delta\varepsilon$ & $\tau$                 & $\alpha$ & $\sigma$               \\
		\midrule
		$S_1$                  & Lunar regolith \cite{carrier1991PhysicalPropertiesLunar,olhoeft1974ElectricalPropertiesLunar}           & 3.0285                 & 0                   & $10^{-12}$             & 0        & $10^{-5}$              \\
		$S_2$                  & Dry basalt/lava \cite{elshafie2013DielectricHardnessMeasurements,ahmad1990MagneticElectricalProperties} & 9.0                    & 0                   & $10^{-12}$             & 0        & $10^{-5}$              \\
		$S_3$                  & Pure ice \cite{auty1952DielectricPropertiesIce,evans1965DielectricPropertiesIce}                        & 3.16                   & 88.34               & $2.1\times10^{-5}$     & 0        & $10^{-5}$              \\
		$S_4$                  & Moist sandy-loam soil \cite{Loewer2017UltraBroadbandSoils,Miller2004RadarDetectionLandmines}            & 21.60                  & 30.49               & $3.60\times10^{-8}$    & 0.45     & $1.95\times10^{-2}$    \\
		$S_{\mathrm{syn}}$     & Synthetic reference                                                                                     & 4.0                    & 2.0                 & $1.0610\times10^{-9}$  & 0.25     & $5\times10^{-3}$       \\
		$S_{\mathrm{balance}}$ & Balanced synthetic case                                                                                 & 5.43374                & 0.110543            & $6.12549\times10^{-6}$ & 0.49     & $6.89221\times10^{-5}$ \\
		\bottomrule
	\end{talltblr}
\end{table*}

\begin{table}[!t]
	\centering
	\caption{Common geometry, FDA, and random-field settings used in all experiments.}
	\label{tab:exp_geometry_setup}
	\begin{talltblr}[
			label = none,
			entry = none,
			remark{Note} = {$f_0$ denotes the first transmit frequency, and the transmit frequencies are generated as $f_n=f_0+(n-1)\Delta f$ for $n=1,\ldots,N$.}
		]
		{
			width=\columnwidth,
			colspec={XXXX},
			row{1}={font=\bfseries}
		}
		\toprule
		Parameter   & Value          & Parameter      & Value              \\
		\midrule
		$N$         & 8              & $M$            & 8                  \\
		$f_0$       & \qty{100}{MHz} & $\Delta f$     & \qty{20}{MHz}      \\
		$d_a$       & \qty{0.05}{m}  & $P$            & 525                \\
		$\Delta x$  & \qty{0.05}{m}  & $\Delta z$     & \qty{0.025}{m}     \\
		$W_y$       & \qty{1.0}{m}   & $\Delta V$     & \qty{1.25e-3}{m^3} \\
		$\ell_x$    & \qty{0.15}{m}  & $\rho_c$       & 0.3                \\
		$w$         & $[1,1,1,1,1]$  & $L$            & 2000               \\
		Random seed & 20260405       & $s_{\mu}$ scan & $[0.0625,4.0]$     \\
		\bottomrule
	\end{talltblr}
\end{table}

\subsection{Validation of the Constitutive and Local Linearized Forward Chain}\label{subsec:exp_chain_verification}\label{subsec:exp_linearization_boundary}

The constitutive layer is first verified by comparing the analytical sensitivity channels $\psi_q$ with their finite-difference references. Across all active channels in all six scenarios, the largest relative derivative error is below $5\times10^{-6}$. This result indicates that the Cole--Cole sensitivity mapping is numerically stable under the present parameter settings, including the lossy soil case $S_4$.

The next question is whether the local first-order chain remains accurate after the constitutive perturbation is inserted into the Born snapshot operator. Table~\ref{tab:exp_local_chain_summary} reports the worst 95th-percentile contrast and snapshot errors over the scanned interval $s_{\mu}\in[0.0625,4]$. No scenario exceeds the prescribed 0.05 threshold within this standardized interval, and no tighter upper bound is identified by the scan under the fixed scaling $\bm D_\mu$. This result should be interpreted as a configuration-specific local-validity diagnostic. The largest closure errors occur in $S_4$, but even there the worst 95th-percentile errors remain below $3.7\times10^{-3}$ for the exact contrast and below $1.9\times10^{-3}$ for the exact Born snapshot. Within the scanned weak-fluctuation range, the constitutive linearization and the linearized Born chain therefore remain numerically consistent.

\begin{table}[!t]
	\centering
	\caption{Validation summary for the constitutive mapping and the local linearized Born chain. The reported contrast and snapshot errors are the worst 95th-percentile values over the full $s_{\mu}$ scan.}
	\label{tab:exp_local_chain_summary}
	\begin{tblr}{
		width=\columnwidth,
		colspec={Q[c,0.85]Q[c,1.0]Q[c,0.9]Q[c,1.05]Q[c,1.0]},
		row{1}={font=\bfseries}
		}
		\toprule
		Scenario               & $\max \epsilon_{\psi_q}$ & Recommended $s_{\mu}$ & Worst $p95(\epsilon_{\xi})$ & Worst $p95(\epsilon_y)$ \\
		\midrule
		$S_1$                  & $<10^{-9}$               & 4.0                   & $1.35\times10^{-5}$         & $7.43\times10^{-6}$     \\
		$S_2$                  & $<10^{-9}$               & 4.0                   & $1.01\times10^{-5}$         & $1.84\times10^{-5}$     \\
		$S_3$                  & $4.98\times10^{-6}$      & 4.0                   & $1.83\times10^{-4}$         & $3.39\times10^{-4}$     \\
		$S_4$                  & $<10^{-5}$               & 4.0                   & $1.83\times10^{-3}$         & $3.67\times10^{-3}$     \\
		$S_{\mathrm{syn}}$     & $<10^{-6}$               & 4.0                   & $5.66\times10^{-4}$         & $4.11\times10^{-4}$     \\
		$S_{\mathrm{balance}}$ & $<10^{-6}$               & 4.0                   & $2.98\times10^{-4}$         & $1.66\times10^{-4}$     \\
		\bottomrule
	\end{tblr}
\end{table}

\subsection{Background-Kernel and FDA-Encoding Effects}\label{subsec:exp_kernel_fda}

The background medium enters the channel model twice: through the constitutive sensitivity $\psi_q(\omega,\bm{x})$ and through the background propagation kernel embedded in $\bm A$. The resulting forward matrices are clearly medium dependent. Among the first three physical scenes, the pairwise discrepancies satisfy $\Delta_A^{(S_1,S_2)}=0.8602$, $\Delta_A^{(S_1,S_3)}=1.6893$, and $\Delta_A^{(S_2,S_3)}=6.8765$. Relative to the free-space kernel, the discrepancies remain between 0.7761 and 1.0471. Under the present setup, the observation operator is therefore not well represented by a free-space surrogate.

The FDA frequency-step scan then shows that the frequency increment is not a passive bookkeeping parameter. Table~\ref{tab:exp_fda_scan} reports representative clutter-structure metrics at $\Delta f\in\{0,20,40\}\,\si{MHz}$, with \qty{20}{MHz} retained as the common operating point used elsewhere in the manuscript. For all four physical scenarios, changing $\Delta f$ alters both the effective rank and the target--clutter overlap. The effect is modest for $S_1$ and $S_3$, stronger for $S_2$, and strongest for the lossy soil case $S_4$, where $\eta_{0.9}$ decreases from 0.7155 at $\Delta f=\qty{0}{MHz}$ to 0.0530 at $\Delta f=\qty{40}{MHz}$. Within the present geometry, frequency range, and representative target-steering setting, FDA encoding therefore changes not only absolute propagation phases but also the normalized covariance geometry used in the structural diagnostics.

\begin{table}[!t]
	\centering
	\caption{Representative FDA scan results for the physical scenes and the representative target-steering setting. The operating point used in the rest of the paper is $\Delta f=\qty{20}{MHz}$.}
	\label{tab:exp_fda_scan}
	\begin{tblr}{
		width=\columnwidth,
		colspec={Q[c,0.8]Q[c,0.8]Q[c,0.8]Q[c,0.8]Q[c,0.8]Q[c,0.8]Q[c,0.8]},
		row{1}={font=\bfseries}
		}
		\toprule
		Scenario & $r_{\mathrm{eff}}(0)$ & $r_{\mathrm{eff}}(20)$ & $r_{\mathrm{eff}}(40)$ & $\eta_{0.9}(0)$ & $\eta_{0.9}(20)$ & $\eta_{0.9}(40)$ \\
		\midrule
		$S_1$    & 2.1538                & 2.5793                 & 2.8847                 & 0.9276          & 0.8210           & 0.5771           \\
		$S_2$    & 2.4596                & 3.0027                 & 3.1368                 & 0.8730          & 0.5611           & 0.3011           \\
		$S_3$    & 2.1624                & 2.5959                 & 2.8968                 & 0.9264          & 0.8138           & 0.5641           \\
		$S_4$    & 2.6477                & 2.7277                 & 2.6244                 & 0.7155          & 0.3146           & 0.0530           \\
		\bottomrule
	\end{tblr}
\end{table}

\subsection{Numerical Validation of the Clutter-Covariance Propagation Formula}\label{subsec:exp_covariance_validation}

The central covariance-propagation relation
\begin{equation*}
	\bm R_c=\bm A\bm R_{\mu}\bm A^{\mathrm H}
\end{equation*}
is validated against both linearized Monte Carlo snapshots and exact-contrast Born snapshots. Table~\ref{tab:exp_covariance_error} shows that the linearized and exact closures are numerically almost indistinguishable at the covariance level: the difference between $\epsilon_{\mathrm{cov}}^{\mathrm{lin}}$ and $\epsilon_{\mathrm{cov}}^{\mathrm{exact}}$ is below $1.2\times10^{-4}$ in all four physical scenes. The exact closure error stays between 0.0221 and 0.0496, while the eigenspectrum and principal-subspace errors remain below 0.0440 and 0.0291, respectively. These values are consistent with finite-sample Monte Carlo discrepancy rather than with a systematic breakdown of the exact-contrast propagation chain.

\begin{table}[!t]
	\centering
	\caption{Monte Carlo closure errors for the theoretical clutter covariance.}
	\label{tab:exp_covariance_error}
	\begin{tblr}{
		width=\columnwidth,
		colspec={Q[c,0.8]Q[c,0.9]Q[c,0.9]Q[c,0.9]Q[c,0.9]},
		row{1}={font=\bfseries}
		}
		\toprule
		Scenario & $\epsilon_{\mathrm{cov}}^{\mathrm{lin}}$ & $\epsilon_{\mathrm{cov}}^{\mathrm{exact}}$ & $\epsilon_{\lambda}^{\mathrm{exact}}$ & $\epsilon_{\mathrm{sub}}^{\mathrm{exact}}$ \\
		\midrule
		$S_1$    & 0.0221                                   & 0.0221                                     & 0.0100                                & 0.0291                                     \\
		$S_2$    & 0.0460                                   & 0.0460                                     & 0.0407                                & 0.0166                                     \\
		$S_3$    & 0.0497                                   & 0.0496                                     & 0.0439                                & 0.0273                                     \\
		$S_4$    & 0.0306                                   & 0.0306                                     & 0.0103                                & 0.0122                                     \\
		\bottomrule
	\end{tblr}
\end{table}

\subsection{Structural Drivers of the Clutter Spectrum and Subspace}\label{subsec:exp_ell}

The first structural factor is the spatial correlation length. In the reference scene $S_2$, increasing $\ell_x$ from \qty{0.05}{m} to \qty{0.40}{m} reduces $r_{\mathrm{eff}}$ from 4.6393 to 1.6883 and contracts $p_{0.9}$ from 4 to 2, while $\eta_{0.9}$ decreases from 0.7639 to 0.3740. Within this scan, the dominant clutter energy becomes progressively more concentrated as the input random field becomes more spatially coherent.

The second structural factor is the background medium itself. Under the same standardized random field, $S_1$ and $S_3$ exhibit strong target--clutter overlap ($\eta_{0.9}>0.81$), $S_2$ is more weakly aligned ($\eta_{0.9}=0.5611$), and $S_4$ yields the smallest baseline overlap ($\eta_{0.9}=0.3146$ and $\gamma_{0.9}=0.6854$). These differences arise even though the same random-field prescription is used in all cases. Under the present model, the background-dependent operator $\bm A$ is therefore a primary structural control variable.

Inter-parameter coupling in $S_{\mathrm{balance}}$ is then used as a negative diagnostic for the physical-scale scans. Varying $\rho_c$ from 0 to 0.9 changes $r_{\mathrm{eff}}$ only from 2.8183 to 2.8173 and shifts $\eta_{0.9}$ only from 0.6959 to 0.6954. The four weighting configurations are similarly close on the physical scale. The block-normalized diagnostic arrays preserve the same qualitative ranking, which indicates that the weak physical-scale sensitivity is mainly associated with the dominance pattern already built into the forward blocks.

\begin{table}[!t]
	\centering
	\caption{Spatial-correlation scan in $S_2$. Larger $\ell_x$ concentrates the clutter spectrum into fewer dominant modes.}
	\label{tab:exp_ell_summary}
	\begin{tblr}{
		width=\columnwidth,
		colspec={Q[c,0.9]Q[c,0.9]Q[c,0.8]Q[c,0.8]Q[c,0.9]Q[c,0.9]},
		row{1}={font=\bfseries}
		}
		\toprule
		$\ell_x$      & $r_{\mathrm{eff}}$ & $p_{0.9}$ & $p_{0.95}$ & $\eta_{0.9}$ & $\gamma_{0.9}$ \\
		\midrule
		\qty{0.05}{m} & 4.6393             & 4         & 6          & 0.7639       & 0.2361         \\
		\qty{0.10}{m} & 3.6243             & 4         & 5          & 0.7611       & 0.2389         \\
		\qty{0.20}{m} & 2.5592             & 3         & 4          & 0.5228       & 0.4772         \\
		\qty{0.40}{m} & 1.6883             & 2         & 2          & 0.3740       & 0.6260         \\
		\bottomrule
	\end{tblr}
\end{table}

\begin{table}[!t]
	\centering
	\caption{Baseline structural metrics across the four physical background scenes.}
	\label{tab:exp_scene_baseline}
	\begin{tblr}{
		width=\columnwidth,
		colspec={Q[c,0.8]Q[c,0.9]Q[c,0.8]Q[c,0.8]Q[c,0.9]Q[c,0.9]Q[c,0.9]},
		row{1}={font=\bfseries}
		}
		\toprule
		Scenario & $r_{\mathrm{eff}}$ & $p_{0.9}$ & $p_{0.95}$ & $\eta_{0.9}$ & $\gamma_{0.9}$ & $\mathrm{tr}(\bm R_c)$ \\
		\midrule
		$S_1$    & 2.5793             & 3         & 4          & 0.8210       & 0.1790         & $1.92\times10^{-6}$    \\
		$S_2$    & 3.0027             & 3         & 4          & 0.5611       & 0.4389         & $5.99\times10^{-7}$    \\
		$S_3$    & 2.5959             & 3         & 4          & 0.8138       & 0.1862         & $7.12\times10^{-7}$    \\
		$S_4$    & 2.7277             & 3         & 4          & 0.3146       & 0.6854         & $5.43\times10^{-8}$    \\
		\bottomrule
	\end{tblr}
\end{table}

\begin{table}[!t]
	\centering
	\caption{Physical-scale correlation and weighting scans in $S_{\mathrm{balance}}$. The induced changes are measurable but secondary compared with the scene and correlation-length effects.}
	\label{tab:exp_corr_weight_summary}
	\begin{tblr}{
		width=\columnwidth,
		colspec={Q[l,1.55]Q[c,0.85]Q[c,0.8]Q[c,0.8]Q[c,0.85]Q[c,0.85]},
		row{1}={font=\bfseries}
		}
		\toprule
		Configuration         & $r_{\mathrm{eff}}$ & $p_{0.9}$ & $p_{0.95}$ & $\eta_{0.9}$ & $\gamma_{0.9}$ \\
		\midrule
		$\rho_c=0.0$          & 2.8183             & 3         & 4          & 0.6959       & 0.3041         \\
		$\rho_c=0.9$          & 2.8173             & 3         & 4          & 0.6954       & 0.3046         \\
		Uniform weighting     & 2.8180             & 3         & 4          & 0.6958       & 0.3042         \\
		Permittivity enhanced & 2.8180             & 3         & 4          & 0.6958       & 0.3042         \\
		Relaxation enhanced   & 2.8170             & 3         & 4          & 0.6953       & 0.3047         \\
		Conductivity enhanced & 2.8181             & 3         & 4          & 0.6956       & 0.3044         \\
		\bottomrule
	\end{tblr}
\end{table}

\subsection{Target-Geometry Robustness}\label{subsec:exp_target_robustness}

The target steering vector is used only as a geometric probe, so its robustness must be tested explicitly. Five target positions are therefore scanned in each physical background scene. Table~\ref{tab:exp_target_robustness} shows that $S_1$ and $S_3$ are both stable and highly overlapping, with mean $\eta_{0.9}$ above 0.82 and standard deviations below 0.10. Scenario $S_2$ remains moderately stable but yields noticeably lower overlap. By contrast, the moist-soil case $S_4$ exhibits both the smallest average overlap and the largest positional variability, with $\eta_{0.9}$ ranging from 0.0211 to 0.6017. Any target-separability statement in this regime should therefore be read as referring to representative steering vectors rather than to a geometry-independent conclusion.

\begin{table}[!t]
	\centering
	\caption{Target-geometry robustness over five representative target points.}
	\label{tab:exp_target_robustness}
	\begin{tblr}{
		width=\columnwidth,
		colspec={Q[c,0.8]Q[c,0.9]Q[c,0.9]Q[c,0.9]Q[c,0.9]},
		row{1}={font=\bfseries}
		}
		\toprule
		Scenario & Mean $\eta_{0.9}$ & Std. $\eta_{0.9}$ & Min $\eta_{0.9}$ & Max $\eta_{0.9}$ \\
		\midrule
		$S_1$    & 0.8320            & 0.0914            & 0.6823           & 0.9343           \\
		$S_2$    & 0.5946            & 0.1760            & 0.3635           & 0.8369           \\
		$S_3$    & 0.8271            & 0.0935            & 0.6754           & 0.9326           \\
		$S_4$    & 0.2844            & 0.2002            & 0.0211           & 0.6017           \\
		\bottomrule
	\end{tblr}
\end{table}

\subsection{Interpretation Boundaries and Overall Summary}\label{subsec:exp_boundary_summary}

Two boundary experiments are used to separate genuine medium-induced structural changes from effects that only alter apparent spectral descriptors. 

First, global covariance scaling $\bm R_c(\kappa)=\kappa \bm R_c$ leaves the normalized spectrum unchanged in both $S_2$ and $S_4$: across $\kappa\in[0.25,4]$, the tuples $(r_{\mathrm{eff}},p_{0.9},\eta_{0.9})$ remain fixed, while $\mathrm{tr}(\bm R_c)$ scales linearly. Pure power amplification is therefore not, by itself, a structural mechanism.

Second, adding a fixed noise floor through $\bm R_y=\bm R_c+\sigma_n^2\bm I$ strongly inflates the spectral tail at low SNR. Table~\ref{tab:exp_noise_summary} shows that, in both $S_2$ and $S_4$, the effective rank and dominant-energy dimension increase sharply at \qty{0}{dB} and then contract toward their noise-free values by \qty{20}{dB}. Low-SNR spectral spreading should therefore be interpreted as a noise-floor effect, not as evidence that the background medium itself has produced a higher-dimensional clutter subspace.

Taken together, the experiments support three conclusions:

\begin{enumerate}
    \item The constitutive-to-covariance propagation chain is numerically consistent over the scanned weak-fluctuation regime.
    \item Spatial correlation length and background-scene variation are consistently strong factors in the tested cases, while the FDA frequency increment also produces measurable changes in the normalized covariance geometry.
    \item Inter-parameter correlation and channel reweighting act only as secondary corrections on the physical scale, whereas global power scaling and a fixed noise floor must be kept separate from genuinely medium-induced subspace reshaping.
\end{enumerate}

\begin{table}[!t]
	\centering
	\caption{Noise-floor boundary experiment for the two interpretation-boundary scenes.}
	\label{tab:exp_noise_summary}
	\begin{tblr}{
		width=\columnwidth,
		colspec={Q[c,0.7]Q[c,0.8]Q[c,0.95]Q[c,0.8]Q[c,0.95]Q[c,0.8]Q[c,0.95]},
		row{1}={font=\bfseries}
		}
		\toprule
		Scene & SNR          & $r_{\mathrm{eff}}$ & $p_{0.9}$ & $p_{0.95}$ & $\eta_{0.9}$ & $\gamma_{0.9}$      \\
		\midrule
		$S_2$ & \qty{0}{dB}  & 24.0579            & 52        & 58         & 1.0000       & $1.22\times10^{-9}$ \\
		$S_2$ & \qty{20}{dB} & 3.2355             & 3         & 5          & 0.5611       & 0.4389              \\
		$S_4$ & \qty{0}{dB}  & 23.1644            & 52        & 58         & 1.0000       & $1.45\times10^{-6}$ \\
		$S_4$ & \qty{20}{dB} & 2.9425             & 3         & 4          & 0.3146       & 0.6854              \\
		\bottomrule
	\end{tblr}
\end{table}

\section{Conclusion}\label{sec:conclusion}

This paper studied medium-induced clutter formation in single-snapshot FDA-MIMO-GPR under Cole--Cole-described subsurface backgrounds. A medium-aware snapshot model was established by linking the Cole--Cole constitutive description, the background Green response, the channel-domain propagation kernels, and the resulting clutter covariance. On this basis, the analysis clarified how weak distributed constitutive fluctuations are transferred into observation-domain spectral changes and target--clutter overlap metrics.

The main value of the present work lies in providing a physically interpretable statistical route from dispersive-medium uncertainty to clutter covariance and subspace structure. The numerical results showed that the constitutive mapping, the background-dependent forward operator, and the covariance-propagation relation are mutually consistent under the adopted setup. They further indicated that spatial correlation length and background-scene variation are consistently strong structural factors, and that the FDA frequency increment can measurably affect the normalized covariance geometry.

The scope of the present theory is intentionally limited. The analysis is restricted to weak distributed non-target medium fluctuations around a prescribed dispersive background. It relies on a background-perturbation formulation, a first-order Born approximation, a first-order constitutive linearization, and a scalarized channel-domain representation. Accordingly, the proposed model should be interpreted as a local weak-fluctuation theory. It is not intended as a full-wave description for strong scattering, strong multiple interactions, or fully vectorial coupling. The target-overlap quantities should also be interpreted as geometry-dependent probes of representative steering vectors, rather than as detection-performance metrics.

The validity of the framework therefore depends on the admissibility of these approximations in the background scenario of interest. Within that range, the proposed formulation provides a useful basis for understanding how weak medium fluctuations reshape clutter covariance and alter dominant-subspace geometry. Beyond that range, higher-order constitutive effects, stronger scattering interactions, and richer channel representations should be incorporated explicitly.

\bibliography{ref}

@article{Loewer2017UltraBroadbandSoils,
  author  = {Loewer, Markus and G{\"u}nther, Thomas and Igel, Jan and Kruschwitz, Sabine and Martin, Tobias and Wagner, Norman},
  title   = {Ultra-broad-band electrical spectroscopy of soils and sediments---a combined permittivity and conductivity model},
  journal = {Geophysical Journal International},
  volume  = {210},
  number  = {3},
  pages   = {1360--1373},
  year    = {2017},
  doi     = {10.1093/gji/ggx242}
}

@article{Miller2004RadarDetectionLandmines,
  author  = {Miller, T. W. and Hendrickx, J. M. H. and Borchers, B.},
  title   = {Radar Detection of Buried Landmines in Field Soils},
  journal = {Vadose Zone Journal},
  volume  = {3},
  number  = {4},
  pages   = {1116--1127},
  year    = {2004},
  doi     = {10.2136/vzj2004.1116}
}

@article{ahmad1990MagneticElectricalProperties,
  title = {Some Magnetic and Electrical Properties of Basalt Rocks},
  author = {Ahmad, M. S. and Zihlif, A. M.},
  date = {1990-11-01},
  journaltitle = {Materials Letters},
  shortjournal = {Mater. Lett.},
  volume = {10},
  number = {4},
  pages = {207--214},
  issn = {0167-577X},
  doi = {10.1016/0167-577X(90)90090-9},
  url = {https://www.sciencedirect.com/science/article/pii/0167577X90900909},
  urldate = {2026-03-26},
  langid = {english}
}

@article{auty1952DielectricPropertiesIce,
  title = {Dielectric Properties of Ice and Solid {{D2O}}},
  author = {Auty, Robert P. and Cole, Robert H.},
  date = {1952-08},
  journaltitle = {Journal of Chemical Physics},
  shortjournal = {J. Chem. Phys.},
  volume = {20},
  number = {8},
  eprint = {https://pubs.aip.org/aip/jcp/article-pdf/20/8/1309/18801220/1309_1_online.pdf},
  pages = {1309--1314},
  issn = {0021-9606},
  doi = {10.1063/1.1700726},
  url = {https://doi.org/10.1063/1.1700726},
  langid = {english}
}

@article{brouet2016CharacterizationPermittivityControlled,
  title = {Characterization of the Permittivity of Controlled Porous Water Ice-Dust Mixtures to Support the Radar Exploration of Icy Bodies},
  author = {Brouet, Y. and Neves, L. and Sabouroux, P. and Levasseur-Regourd, A. C. and Poch, O. and Encrenaz, P. and Pommerol, A. and Thomas, N. and Kofman, W.},
  date = {2016},
  journaltitle = {Journal of Geophysical Research: Planets},
  shortjournal = {J. Geophys. Res.: Planets},
  volume = {121},
  number = {12},
  pages = {2426--2443},
  issn = {2169-9100},
  doi = {10.1002/2016JE005045},
  url = {https://onlinelibrary.wiley.com/doi/abs/10.1002/2016JE005045},
  urldate = {2026-03-26},
  langid = {english}
}

@incollection{carrier1991PhysicalPropertiesLunar,
  title = {Physical Properties of the Lunar Surface},
  booktitle = {Lunar Sourcebook, a User's Guide to the Moon},
  author = {Carrier, Iii, W. D. and Olhoeft, G. R. and Mendell, W.},
  date = {1991-01-01},
  pages = {475--594},
  url = {https://ui.adsabs.harvard.edu/abs/1991lsug.book..475C},
  urldate = {2026-03-26},
  langid = {english}
}

@article{cole1941DispersionAbsorptionDielectrics,
  title = {Dispersion and Absorption in Dielectrics {{I}}. {{Alternating}} Current Characteristics},
  author = {Cole, Kenneth S. and Cole, Robert H.},
  date = {1941-04},
  journaltitle = {Journal of Chemical Physics},
  shortjournal = {J. Chem. Phys.},
  volume = {9},
  number = {4},
  eprint = {https://pubs.aip.org/aip/jcp/article-pdf/9/4/341/18792177/341_1_online.pdf},
  pages = {341--351},
  issn = {0021-9606},
  doi = {10.1063/1.1750906},
  url = {https://doi.org/10.1063/1.1750906},
  langid = {american}
}

@article{elshafie2013DielectricHardnessMeasurements,
  title = {Dielectric and Hardness Measurements of Planetary Analog Rocks in Support of In-Situ Subsurface Sampling},
  author = {ElShafie, Ahmed and Heggy, Essam},
  date = {2013},
  journaltitle = {Planetary and Space Science},
  shortjournal = {Planet. Space Sci.},
  volume = {86},
  pages = {150--154},
  issn = {0032-0633},
  doi = {10.1016/j.pss.2013.02.003},
  url = {https://www.sciencedirect.com/science/article/pii/S0032063313000329},
  langid = {english}
}

@article{evans1965DielectricPropertiesIce,
  title = {Dielectric Properties of Ice and Snow–a Review},
  author = {Evans, S.},
  date = {1965-01},
  journaltitle = {Journal of Glaciology},
  shortjournal = {J. Glaciol.},
  volume = {5},
  number = {42},
  pages = {773--792},
  issn = {0022-1430, 1727-5652},
  doi = {10.3189/S0022143000018840},
  url = {https://www.cambridge.org/core/journals/journal-of-glaciology/article/dielectric-properties-of-ice-and-snowa-review/A4F950358E513838DAE111F6A07EC077},
  urldate = {2026-03-26},
  langid = {english}
}

@article{feng2022ShallowRegolithStructure,
  title = {Shallow Regolith Structure and Obstructions Detected by Lunar Regolith Penetrating Radar at Chang’{{E-5}} Drilling Site},
  author = {Feng, Jianqing and Siegler, Matthew A. and White, Mackenzie N.},
  date = {2022-01},
  journaltitle = {Remote Sensing},
  shortjournal = {Remote Sens.},
  volume = {14},
  number = {14},
  pages = {3378},
  publisher = {Multidisciplinary Digital Publishing Institute},
  issn = {2072-4292},
  doi = {10.3390/rs14143378},
  url = {https://www.mdpi.com/2072-4292/14/14/3378},
  urldate = {2026-04-03},
  langid = {english}
}

@article{gennarelli2026EffectEquivalentPermittivity,
  title = {The Effect of the Equivalent Permittivity Model in Contactless {{MIMO-GPR}} Imaging},
  author = {Gennarelli, Gianluca and Catapano, Ilaria and Soldovieri, Francesco},
  date = {2026-02-26},
  journaltitle = {Sensors},
  shortjournal = {Sens.},
  volume = {26},
  number = {5},
  publisher = {Multidisciplinary Digital Publishing Institute},
  issn = {1424-8220},
  doi = {10.3390/s26051463},
  url = {https://www.mdpi.com/1424-8220/26/5/1463},
  urldate = {2026-03-28},
  langid = {english}
}

@article{kumar2025EnhancingSubsurfaceExploration,
  title = {Enhancing Subsurface Exploration: A Comprehensive Review of Advanced Clutter Removal Techniques for Ground Penetrating Radar Imaging},
  shorttitle = {Enhancing Subsurface Exploration},
  author = {Kumar, Buddepu Santhosh and Baraha, Satyakam and Sahoo, Ajit Kumar and Maiti, Subrata},
  date = {2025-01-15},
  journaltitle = {Measurement},
  shortjournal = {Measurement},
  volume = {239},
  pages = {115432},
  issn = {0263-2241},
  doi = {10.1016/j.measurement.2024.115432},
  url = {https://www.sciencedirect.com/science/article/pii/S0263224124013174},
  urldate = {2026-04-03},
  langid = {english}
}

@article{li2026AdaptiveTargetDetection,
  title = {Adaptive Target Detection for {{FDA-MIMO}} Radar with Training Data in Gaussian Noise},
  author = {Li, Ping and Huang, Bang and Wang, Wen-Qin},
  date = {2026},
  journaltitle = {IEEE Transactions on Vehicular Technology},
  shortjournal = {IEEE Trans. Veh. Technol.},
  pages = {1--16},
  issn = {1939-9359},
  doi = {10.1109/TVT.2026.3662208},
  url = {https://ieeexplore.ieee.org/document/11373410},
  urldate = {2026-03-28},
  langid = {english}
}

@article{liu2018DetectionSubsurfaceTarget,
  title = {Detection of Subsurface Target Based on {{FDA-MIMO}} Radar},
  author = {Liu, Qinghua and Jiang, Chang and Jin, Liangnian and Ouyang, Shan},
  date = {2018},
  journaltitle = {International Journal of Antennas and Propagation},
  shortjournal = {Int. J. Antennas Propag.},
  volume = {2018},
  number = {1},
  pages = {8629806},
  issn = {1687-5877},
  doi = {10.1155/2018/8629806},
  url = {https://onlinelibrary.wiley.com/doi/abs/10.1155/2018/8629806},
  urldate = {2026-03-28},
  langid = {english}
}

@article{masoodi2024MultiviewMultistaticVs,
  title = {Multiview Multistatic vs. {{Multimonostatic}} Three-Dimensional Ground-Penetrating Radar Imaging: A Comparison},
  shorttitle = {Multiview Multistatic vs. {{Multimonostatic}} Three-Dimensional Ground-Penetrating Radar Imaging},
  author = {Masoodi, Mehdi and Gennarelli, Gianluca and Soldovieri, Francesco and Catapano, Ilaria},
  date = {2024-08-27},
  journaltitle = {Remote Sensing},
  shortjournal = {Remote Sens.},
  volume = {16},
  number = {17},
  publisher = {Multidisciplinary Digital Publishing Institute},
  issn = {2072-4292},
  doi = {10.3390/rs16173163},
  url = {https://www.mdpi.com/2072-4292/16/17/3163},
  urldate = {2026-03-28},
  langid = {english}
}

@article{olhoeft1973ElectricalPropertiesLunar,
  title = {Electrical Properties of Lunar Solid Samples},
  author = {Olhoeft, Gary and Frisillo, A. and Strangway, D. and Sharpe, Howard},
  date = {1973-02},
  volume = {4},
  pages = {575},
  langid = {english}
}

@article{olhoeft1974ElectricalPropertiesLunar,
  title = {Electrical Properties of Lunar Soil Sample 15301,38},
  author = {Olhoeft, G. R. and Frisillo, A. L. and Strangway, D. W.},
  date = {1974},
  journaltitle = {Journal of Geophysical Research (1896-1977)},
  shortjournal = {J. Geophys. Res. (1896-1977)},
  volume = {79},
  number = {11},
  pages = {1599--1604},
  issn = {2156-2202},
  doi = {10.1029/JB079i011p01599},
  url = {https://onlinelibrary.wiley.com/doi/abs/10.1029/JB079i011p01599},
  urldate = {2026-03-26},
  langid = {english}
}

@article{qin2023FullwaveformInversionGroundpenetrating,
  title = {Full-Waveform Inversion of Ground-Penetrating Radar Data in Frequency-Dependent Media Involving Permittivity Attenuation},
  author = {Qin, Tan and Bohlen, Thomas and Allroggen, Niklas},
  date = {2023-01-01},
  journaltitle = {Geophysical Journal International},
  shortjournal = {Geophys. J. Int.},
  volume = {232},
  number = {1},
  pages = {504--522},
  issn = {0956-540X},
  doi = {10.1093/gji/ggac319},
  url = {https://doi.org/10.1093/gji/ggac319},
  urldate = {2026-03-28}
}

@article{salinasnaval2018GPRClutterAmplitude,
  title = {{{GPR}} Clutter Amplitude Processing to Detect Shallow Geological Targets},
  author = {Salinas Naval, Victor and Santos-Assunçao, Sonia and Pérez-Gracia, Vega},
  date = {2018-01},
  journaltitle = {Remote Sensing},
  shortjournal = {Remote Sens.},
  volume = {10},
  number = {1},
  pages = {88},
  publisher = {Multidisciplinary Digital Publishing Institute},
  issn = {2072-4292},
  doi = {10.3390/rs10010088},
  url = {https://www.mdpi.com/2072-4292/10/1/88},
  urldate = {2026-04-03},
  langid = {english}
}

@article{strangway1972ElectricalPropertiesLunar,
  title = {Electrical Properties of Lunar Soil Dependence on Frequency, Temperature and Moisture},
  author = {Strangway, D.W. and Chapman, W.B. and Olhoeft, G.R. and Carnes, J.},
  date = {1972},
  journaltitle = {Earth and Planetary Science Letters},
  shortjournal = {Earth Planet. Sci. Lett.},
  volume = {16},
  number = {2},
  pages = {275--281},
  issn = {0012-821X},
  doi = {10.1016/0012-821X(72)90203-8},
  url = {https://www.sciencedirect.com/science/article/pii/0012821X72902038},
  langid = {english}
}

@inproceedings{takahashi2010InfluenceSoilInhomogeneity,
  title = {Influence of Soil Inhomogeneity on {{GPR}} for Landmine Detection},
  booktitle = {Proceedings of the {{XIII International Conference}} on {{Ground Penetrating Radar}}},
  author = {Takahashi, Kazunori and Igel, Jan and Preetz, Holger},
  date = {2010-06},
  pages = {1--6},
  doi = {10.1109/ICGPR.2010.5550150},
  url = {https://ieeexplore.ieee.org/document/5550150},
  urldate = {2026-04-03},
  eventtitle = {The {{XIII Internarional Conference}} on {{Ground Penetrating Radar}}},
  langid = {english}
}

@article{takahashi2012ModelingGPRClutter,
  title = {Modeling of {{GPR}} Clutter Caused by Soil Heterogeneity},
  author = {Takahashi, Kazunori and Igel, Jan and Preetz, Holger},
  date = {2012},
  journaltitle = {International Journal of Antennas and Propagation : IJAP},
  shortjournal = {Int. J. Antennas Propag. : IJAP},
  volume = {2012},
  pages = {643430},
  publisher = {New York, NY : Hindawi},
  doi = {10.34657/7178},
  url = {https://oa.tib.eu/renate/handle/123456789/8138},
  urldate = {2026-04-02},
  langid = {english}
}

@article{wang2022RangeAmbiguousClutterSuppression,
  title = {Range-{{Ambiguous Clutter Suppression}} via {{FDA MIMO Planar Array Radar}} with {{Compressed Sensing}}},
  author = {Wang, Yuzhuo and Zhu, Shengqi and Lan, Lan and Li, Ximin and Liu, Zhixin and Wu, Zhixia},
  date = {2022-04-15},
  journaltitle = {Remote Sensing},
  shortjournal = {Remote Sens.},
  volume = {14},
  number = {8},
  publisher = {Multidisciplinary Digital Publishing Institute},
  issn = {2072-4292},
  doi = {10.3390/rs14081926},
  url = {https://www.mdpi.com/2072-4292/14/8/1926},
  urldate = {2026-03-28},
  langid = {english}
}

@article{xie2021GPRUncertaintyModelling,
  title = {{{GPR}} Uncertainty Modelling and Analysis of Object Depth Based on Constrained Least Squares},
  author = {Xie, Fei and Lai, Wallace W. L. and Dérobert, Xavier},
  date = {2021-10-01},
  journaltitle = {Measurement},
  shortjournal = {Measurement},
  volume = {183},
  pages = {109799},
  issn = {0263-2241},
  doi = {10.1016/j.measurement.2021.109799},
  url = {https://www.sciencedirect.com/science/article/pii/S026322412100751X},
  urldate = {2026-04-02},
  langid = {english}
}

@article{zhang2025SubsoilStructureChangE6,
  title = {Subsoil Structure at the Chang’{{E-6}} Landing Site Revealed by in-Situ Lunar Regolith Penetrating Radar},
  author = {Zhang, Zongyu and Ding, Chunyu and Su, Yan and Shen, Shaoxiang and Lu, Wei and Soldovieri, Francesco and Xiao, Zhiyong and Zeng, Xingguo and Du, Wei and Liu, Yuhang and Jiang, Changzhi and Gennarelli, Gianluca and Catapano, Ilaria and Dai, Shun and Feng, Jianqing and Wang, Yichen and Liu, Jianjun and Li, Chunlai},
  date = {2025-08-07},
  journaltitle = {Communications Earth \& Environment},
  shortjournal = {Commun. Earth Environ.},
  volume = {6},
  number = {1},
  pages = {640},
  publisher = {Nature Publishing Group},
  issn = {2662-4435},
  doi = {10.1038/s43247-025-02631-4},
  url = {https://www.nature.com/articles/s43247-025-02631-4},
  urldate = {2026-04-03},
  langid = {english}
}

@article{holm2020TimeDomainCharacterization,
  title = {Time Domain Characterization of the Cole-Cole Dielectric Model},
  author = {Holm, Sverre},
  date = {2020-12-31},
  journaltitle = {Journal of Electrical Bioimpedance},
  shortjournal = {J. Electr. Bioimpedance},
  volume = {11},
  number = {1},
  eprint = {33584910},
  eprinttype = {pubmed},
  pages = {101--105},
  issn = {1891-5469},
  doi = {10.2478/joeb-2020-0015},
  url = {https://pmc.ncbi.nlm.nih.gov/articles/PMC7851980/},
  urldate = {2026-03-14},
  langid = {english},
  pmcid = {PMC7851980}
}

@article{jia2026FDAMIMORadarParameter,
  title = {{{FDA-MIMO}} Radar Parameter Designing against Range-Ambiguous Clutter and Scatter-Wave Jamming},
  author = {Jia, Mingjie and Sun, Yan and Wang, Wen-Qin},
  date = {2026},
  journaltitle = {Signal Processing},
  shortjournal = {Signal Process.},
  volume = {240},
  pages = {110361},
  issn = {0165-1684},
  doi = {10.1016/j.sigpro.2025.110361},
  url = {https://www.sciencedirect.com/science/article/pii/S0165168425004773},
  langid = {english}
}

@online{liu2016ClutterRanksFrequency,
  title = {On Clutter Ranks of Frequency Diverse Radar Waveforms},
  author = {Liu, Yimin and Xiao, Le and Wang, Xiqin and Nehorai, Arye},
  date = {2016-03-27},
  eprint = {1603.08189},
  eprinttype = {arXiv},
  eprintclass = {cs},
  doi = {10.48550/arXiv.1603.08189},
  url = {http://arxiv.org/abs/1603.08189},
  urldate = {2026-03-14},
  langid = {english},
  pubstate = {prepublished}
}

@inproceedings{majchrowska2021ModellingArbitraryComplex,
  title = {Modelling Arbitrary Complex Dielectric Properties – an Automated Implementation for {{gprMax}}},
  booktitle = {2021 11th International Workshop on Advanced Ground Penetrating Radar ({{IWAGPR}})},
  author = {Majchrowska, Sylwia and Giannakis, Iraklis and Warren, Craig and Giannopoulos, Antonios},
  date = {2021-12},
  pages = {1--5},
  issn = {2687-7899},
  doi = {10.1109/IWAGPR50767.2021.9843152},
  langid = {english}
}

@article{oliveira2021GPRClutterReflection,
  title = {{{GPR}} Clutter Reflection Noise-Filtering through Singular Value Decomposition in the Bidimensional Spectral Domain},
  author = {Oliveira, Rui Jorge and Caldeira, Bento and Teixidó, Teresa and Borges, José Fernando},
  date = {2021},
  journaltitle = {Remote Sensing},
  shortjournal = {Remote Sens.},
  volume = {13},
  number = {2005},
  issn = {2072-4292},
  doi = {10.3390/rs13102005},
  url = {https://www.mdpi.com/2072-4292/13/10/2005},
  issue = {10},
  langid = {american}
}

@article{sun2024SpaceTimeRange,
  title = {Space–Time-Range Clutter Suppression via Tensor-Based {{STAP}} for Airborne {{FDA-MIMO}} Radar},
  author = {Sun, Yan and Wang, Wen-Qin and Jiang, Chen},
  date = {2024},
  journaltitle = {Signal Processing},
  shortjournal = {Signal Process.},
  volume = {214},
  pages = {109235},
  issn = {0165-1684},
  doi = {10.1016/j.sigpro.2023.109235},
  url = {https://www.sciencedirect.com/science/article/pii/S0165168423003092},
  langid = {english}
}

@article{wang2022ClutterRankAnalysis,
  title = {Clutter Rank Analysis in Airborne {{FDA-MIMO}} Radar with Range Ambiguity},
  author = {Wang, Keyi and Liao, Guisheng and Xu, Jingwei and Zhang, Yuhong and Huang, Lei},
  date = {2022-04},
  journaltitle = {IEEE Transactions on Aerospace and Electronic Systems},
  shortjournal = {IEEE Trans. Aerosp. Electron. Syst.},
  volume = {58},
  number = {2},
  pages = {1416--1430},
  issn = {1557-9603},
  doi = {10.1109/TAES.2021.3122822},
  langid = {english}
}

@article{WEN2019280,
  title = {Clutter Suppression for Airborne {{FDA-MIMO}} Radar Using Multi-Waveform Adaptive Processing and Auxiliary Channel {{STAP}}},
  author = {Wen, Cai and Tao, Mingliang and Peng, Jinye and Wu, Jianxin and Wang, Tong},
  date = {2019},
  journaltitle = {Signal Processing},
  shortjournal = {Signal Process.},
  volume = {154},
  pages = {280--293},
  issn = {0165-1684},
  doi = {10.1016/j.sigpro.2018.09.016},
  url = {https://www.sciencedirect.com/science/article/pii/S0165168418303001},
  langid = {english}
}

@article{worthmann2021ClutterDistributionsTomographic,
  title = {Clutter Distributions for Tomographic Image Standardization in Ground-Penetrating Radar},
  author = {Worthmann, Brian M. and Chambers, David H. and Perlmutter, David S. and Mast, Jeffrey E. and Paglieroni, David W. and Pechard, Christian T. and Stevenson, Garrett A. and Bond, Steven W.},
  date = {2021-09},
  journaltitle = {IEEE Transactions on Geoscience and Remote Sensing},
  shortjournal = {IEEE Trans. Geosci. Remote Sens.},
  volume = {59},
  number = {9},
  pages = {7957--7967},
  issn = {1558-0644},
  doi = {10.1109/TGRS.2021.3051566},
  url = {https://ieeexplore.ieee.org/abstract/document/9336275},
  urldate = {2025-04-23},
  langid = {american}
}
\bibliographystyle{IEEEtran}

\end{document}